\documentclass[a4,11pt]{article}
\usepackage{authblk}
\usepackage{graphicx}
\title{Observation of coherent two-photon emission from the first vibrationally-excited state of hydrogen molecules}

\author[1]{Yuki Miyamoto}
\author[1]{Hideaki Hara}
\author[1]{Susumu Kuma\thanks{Present address: Atomic, Molecular and Optical Physics Laboratory, RIKEN, Wako, Saitama 351-0198, Japan}}
\author[1]{Takahiko Masuda}
\author[1]{Itsuo Nakano}
\author[2]{Chiaki Ohae\thanks{Present address: Department of Engineering Science, University of Electro-Communications, Chofu, Tokyo 182-8585, Japan}}
\author[1]{Noboru Sasao\thanks{Email:sasao@okayama-u.ac.jp}}
\author[3]{Minoru Tanaka}
\author[4]{Satoshi Uetake\thanks{Email:uetake@okayama-u.ac.jp}}
\author[1]{Akihiro Yoshimi}
\author[1]{Koji Yoshimura}
\author[4]{Motohiko Yoshimura}
\affil[1]{Research Core for Extreme Quantum World, Okayama University, Okayama 700-8530, Japan}
\affil[2]{Graduate School of Natural Science and Technology, Okayama University, Okayama 700-8530, Japan}
\affil[3]{Department of Physics, Osaka University, Toyonaka, Osaka 560-0045, Japan}
\affil[4]{Research Center of Quantum Universe, Okayama University, Okayama 700-8530, Japan}
\date{}
\begin{document}

\maketitle
\begin{abstract}
In this paper, we describe an experiment which was conducted to explore the macro-coherent amplification mechanism 
using a two-photon emission process from the first vibrationally-excited state of para-hydrogen molecule.
Large coherence in the initial state was prepared by the adiabatic Raman method, and 
the lowest Stokes sideband was used as a trigger field.
We observed the coherent two-photon emission consistent with the expectation of the Maxwell-Bloch equation derived for the process,
whose rate is larger by many orders of magnitude than that of the spontaneous emission.
\end{abstract}

\section{Introduction}  
Coherence among an ensemble of atoms or molecules mediated by radiation fields has shown a variety of remarkable phenomena, 
and has offered a platform of devising new tools and/or methods.
One classical example of such coherence is super-radiance\cite{Dicke}\cite{SR-review}. 
In this case, excited atoms or molecules organize themselves into a coherent state via a series of spontaneous emission, 
eventually resulting in explosive radiation pulses.
Another example is the adiabatic Raman process studied by Ref. \cite{Harris}\cite{Fam}\cite{AVSokolov:PRL2000}\cite{Hakuda}.
In this case, the coherence is used to generate a series of equally-spaced sidebands 
which in turn enables one to create ultra-short pulses.

Recently, some of the authors have proposed to use a new type of coherent amplification mechanism in order to study experimentally 
 much suppressed processes involving neutrinos \cite{PTEP}\cite{Macro-Coherence-1}\cite{Dinh}\cite{Core-RENP}.
The ultimate goal of the proposal is to investigate unknown neutrino properties such as 
	their absolute masses, mass type (Dirac or Majorana), and CP-violating phases\cite{PTEP}\cite{Dinh}\cite{Core-RENP}.
This amplification by coherence, termed as macro-coherent amplification, is applicable to a
	process which emits plural particles.
If outgoing particles satisfy a certain phase matching condition, equivalent to the momentum conservation law,
then the process rate becomes proportional to $N^2$, where $N$ is the number of coherent atoms or molecules involved in the process.
When the macro-coherent amplification is applied to two-photon emission process, a pair of intense radiations may emerge 
in a similar fashion to the triggered super-radiance\cite{PSRvsSR}. Such a process, called paired super-radiance (PSR in short), 
has been predicted, and its master equations have been derived \cite{PSR-dynamics}\cite{PTEP}.
In an ideal situation, most of the energy stored in an upper level may be released in an explosive way.
The theory of PSR also predicts much milder events in which the degree of coherence, target number density, decoherence time, 
or a combination of these is less favorable than the explosive one. 

\begin{figure}[h]
	\begin{center}
		\includegraphics[width=0.5\textwidth]{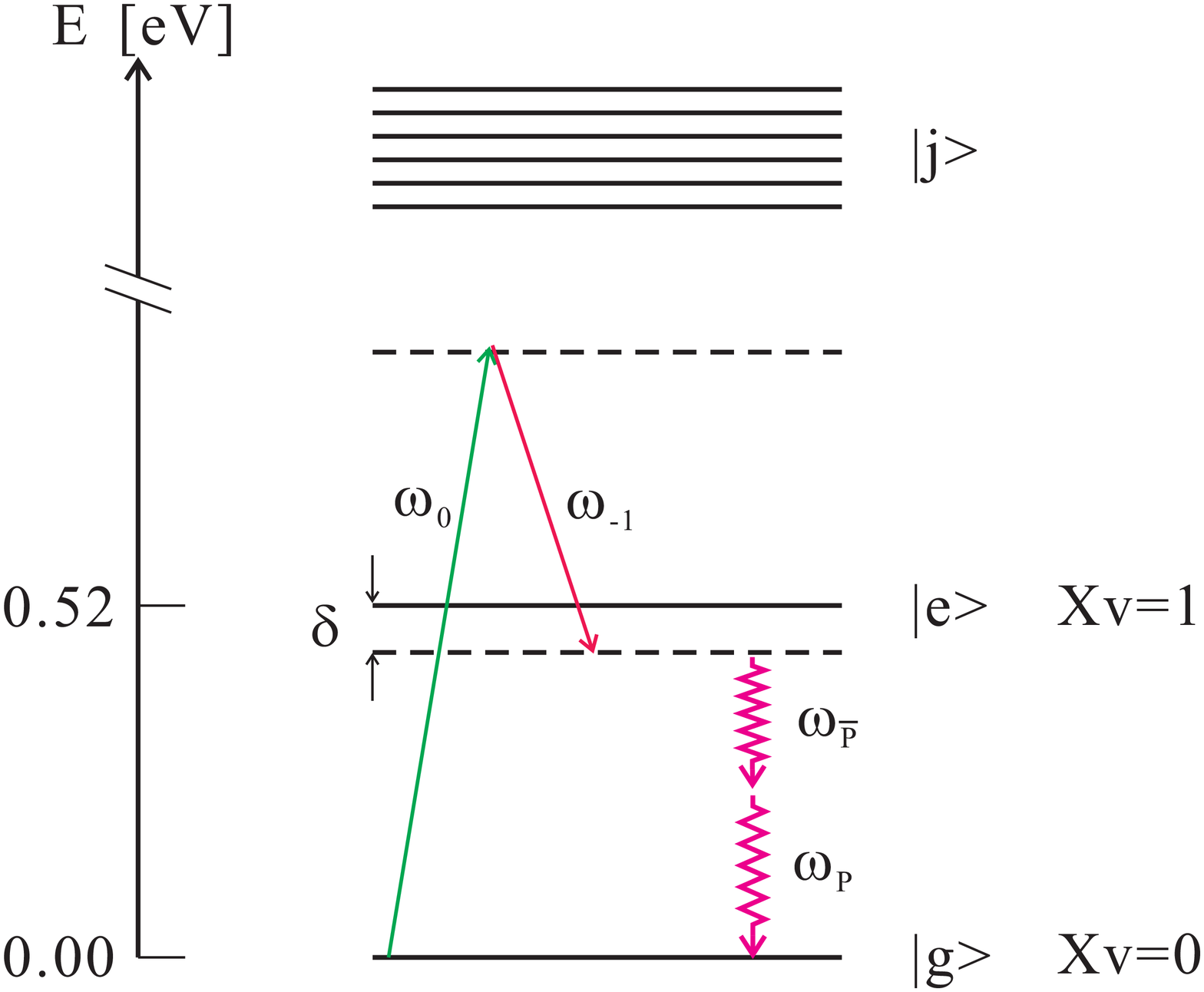}
        \end{center}
	\caption{Schematic diagram showing the relevant hydrogen molecule energy levels and the Raman excitation and two-photon emission processes.}
	\label{fig:H2 Energy Levels}
\end{figure}

In this paper, we describe an experiment which was conducted to explore the macro-coherent amplification mechanism 
using the two-photon emission process from electronically-ground vibrationally-excited state ($|e \rangle$; $Xv=1$)
 of hydrogen molecules.
(For brevity the word ``hydrogen" may be used for hydrogen molecule below.) 
Figure \ref{fig:H2 Energy Levels} shows the hydrogen energy levels relevant to the present experiment.
To prepare the initial states, we employed an adiabatic Raman method 
	changing the hydrogen from its ground ($|g \rangle; Xv=0$) state
	to the superposed state of $|g \rangle$ and $|e \rangle$
	by a pair of driving lasers ($\omega_{0}$ and $\omega_{-1}$ in Fig.\ref{fig:H2 Energy Levels}).
Since the electric dipole (E1) transition is strictly forbidden 
	from electronically-ground vibrationally-excited states of homo-nuclear diatomic molecules, 
        deexcitation from $Xv=1$ is via the two-photon emission whose spontaneous rate is very small.
Two-photon emissions ($\omega_{p}$ and $\omega_{\overline{p}}$ in Fig.\ref{fig:H2 Energy Levels}) 
together with other Raman sidebands were detected in the present experiment, 
and their yields were compared with the theoretical expectations.

In the present work, the two driving lasers were injected in the same direction, 
	and one of the generated sidebands was used as a trigger field.
This experimental configuration has been discussed in the literature
\cite{Parametric-Down-Conversion} in the context of the parametric down conversion.
Our experimental results may be understood from its view point; however, we employ a different approach in explanations below, 
namely the view point of the adiabatic Raman excitation supplemented by the paired super-radiance\cite{PSRvsPDC}.
The basic equation (Maxwell-Bloch) presented below is derived from this view point\cite{PDC2}.

The rest of the paper is organized as follows.
In the next section, we briefly describe theoretical aspects of the paired super-radiance and adiabatic Raman process, 
and present a simulation method based on an effective Hamiltonian combined for both. 
They are non-linear processes and thus demand numerical simulations 
to obtain various observables which can be compared directly with actual experimental data.
Following these, we describe our experimental setup in Sec.3.
The results and conclusions are given in Sec.4 and 5, respectively.

\section{Theory and Simulation}  
We begin our discussion by constructing an effective Hamiltonian which describes both two-photon emission and Raman excitation processes.
The basic QED interaction is the electric dipole interaction (E1) represented by $-\vec{d}\cdot \vec{E}$ with $\vec{d}$ being the dipole moment and $\vec{E}$ 
electric fields.
(We will omit the vector notation below since all the fields treated in this paper are linearly polarized in the same direction.)
In the present system, the E1 dipole interaction connects $|g\rangle$ and $|e\rangle$ through an intermediate state $|j \rangle$, 
which is taken as an electronically-excited state. 
Many intermediate levels may contribute, as shown in Fig. \ref{fig:H2 Energy Levels}, 
	but in the following we consider only one for simplicity.
Extension to the case of multi levels is trivial, and our actual simulation includes several tens of intermediate states \cite{Fam}.
The present system can be regarded as a two level system once the intermediate state $|j \rangle$ is integrated out 
from the Schr{\"o}dinger equation with the aid of the Markov approximation.
The electromagnetic fields to be considered are the two driving lasers  
and the associated Raman sidebands with frequencies of 
\begin{equation}
	\omega_{q}=\omega_{0}+q \Delta \omega, \qquad \Delta \omega=\omega_{0}-\omega_{-1}, 
        \label{Raman sideband frequency}
\end{equation}
where the Raman order $q$ is a positive (anti-Stokes) or negative (Stokes) integer satisfying $\omega_{q}>0$.
In the present experimental conditions, the smallest $q$ (the lowest Stokes sideband) is $q=-4$.
The frequency difference of the two driving lasers $\Delta \omega$ should be chosen to be nearly on resonance; 
the difference between the exact resonance frequency $\omega_{eg}\equiv \omega_{e}-\omega_{g}$ and $\Delta \omega$ is the detuning $\delta \equiv \omega_{eg}- \Delta \omega$.
In addition to these sidebands, there should be the fields corresponding to the two-photon emissions.
The frequencies of the pair are denoted by $\omega_{p}$ and $\omega_{\overline{p}}$ which should satisfy 
the energy conservation law of 
\begin{equation}
	\omega_{p}+\omega_{\overline{p}}=\Delta \omega.
        \label{two-photon cascade frequency}
\end{equation}
Since the Raman excitation imprints a spatially dependent phase of $e^{i \Delta \omega \cdot x/c}$ in medium, 
	eq.(\ref{two-photon cascade frequency}) satisfies the momentum conservation law if the two-photon fields propagate 
        in the same direction as the excitation fields, 
        and the dispersion in the medium is negligible for these wavelengths. 
The macro-coherent amplification mechanism requires the momentum conservation as well as the energy conservation of elementary process.
All the electromagnetic fields are taken to be traveling in one direction (taken to be $x$), and are expressed by
\begin{equation}
	\widetilde{E}_{m}(t,x)=\frac{1}{2}{E}_{m}(t,x)e^{-i \omega_{m}(t-x/c)}+c.c.
        \label{electromagnetic field}
\end{equation}
where $m$ denotes either $p$, $\overline{p}$ or $q$, and $E_{m}$ is the slowly varying envelope function.
For future reference, we list important frequencies in terms of wavelength in Table \ref{wavelength table}.
\begin{table}[h]
\caption{Wavelengths important in the present experiment.}
\begin{center}
\begin{tabular}{rrcc} \hline
name    & wavelength [nm]  & symbols & remark \\ \hline
Energy gap ($|e\rangle - |g\rangle$)& 2403.172 & $\omega_{eg}$ & Ref.\cite{Xv1-energy} \\
Driving laser ($\omega_{0}$)  & 532.216 & $\omega_{0}$ & measured\\
Driving laser ($\omega_{-1}$) & 683.610 & $\omega_{-1}$ & measured\\
Lowest Raman sideband  & 4662.48 & $\omega_{-4}=\omega_{\overline{p}}$ & calculated\\
Two-photon partner & 4959.43 & $\omega_{p}$ & calculated \\ \hline
\end{tabular}
\label{wavelength table}
\end{center}
\vspace{2mm}
Note: All the values are at the pressure of 60 kPa except $\omega_{eg}$ which is the value of the zero pressure (collision-less) limit.
\end{table}

In order to proceed further, we resort to the standard technique of rotating wave (RWA) and slowly varying envelope approximations (SVEA)\cite{RWA-SVEA}. 
When these are applied to the Schr{\"o}dinger equation, it turns out that 
the Raman process as well as the two-photon emission are described by the Hamiltonian ${H}={H}_{0}+{H}_{1}+{H}_{2}$ 
of the form, 
\begin{eqnarray}
  &&{H}_{0}=- \frac{1}{4} \sum_{m=p,\overline{p},q}\left( \begin{array}{cc} \varepsilon_{0}\,\alpha_{gg}^{(m)}|E_{m}|^2  & 0 \\ 0 &  \varepsilon_{0}\,\alpha_{ee}^{(m)}|E_{m}|^2 -4 \hbar \delta \end{array} \right), 
  \label{hamiltonian-0} \\ 
  &&{H}_{1}=-\frac{1}{4} \sum_{q=-4}^{\infty} \left( \begin{array}{cc} 0 & \varepsilon_{0}\,\alpha_{ge}^{(q)}E_{q}E_{q+1}^{\ast}  \\ \varepsilon_{0}\,\alpha_{eg}^{(q)} E_{q}^{\ast}E_{q+1}  & 0 \end{array} \right),
  \label{hamiltonian-1} \\ 
  &&{H}_{2}=-\frac{1}{4} \left( \begin{array}{cc} 0 & \varepsilon_{0}\,\alpha_{ge}^{(p\overline{p})} E_{p}^{\ast}E_{\overline{p}}^{\ast} \\ \varepsilon_{0}\,\alpha_{eg}^{(p\overline{p})} E_{p}E_{\overline{p}}  & 0 \end{array} \right),
  \label{hamiltonian-2}
\end{eqnarray}
where ${H}_{0}$ gives the Stark energy shift with the detuning $\delta$, 
${H}_{1}$ the adiabatic Raman process derived in \cite{Harris}\cite{Hakuda}, 
and ${H}_{2}$ the two-photon emission  which can be reduced from the paired super-radiance master equation when 
electromagnetic propagations are assumed uni-directional \cite{PSR-dynamics}\cite{PTEP}.
In eq.(\ref{hamiltonian-0}), $\alpha_{gg}^{(m)}$ or $\alpha_{ee}^{(m)}$ is 
the polarizability of the state $|g\rangle$ or $|e\rangle$, and is given by \cite{polarizability-note}
\begin{eqnarray}
  &&\alpha_{aa}^{(m)}=\frac{|d_{aj}|^2}{ \varepsilon_{0} \hbar}\left( \frac{1}{\omega_{ja}+\omega_{m}}+\frac{1}{\omega_{ja}-\omega_{m}} \right),
  \qquad (a=g,e; \ m=p, \overline{p},q)
  \label{polarizability-diagonal}
\end{eqnarray}
where $d_{aj}$ and $\hbar \omega_{ja}\equiv \hbar(\omega_{j}-\omega_{a})$ are, respectively, a transition dipole moment 
and energy difference between levels $a-j$.
Similarly the off-diagonal parts of the polarizability in eq.(\ref{hamiltonian-1}-\ref{hamiltonian-2}) are given by 
\begin{eqnarray}
  &&\alpha_{ge}^{(q)}=\alpha_{eg}^{(q)}=\frac{d_{gj}d_{je}}{ \varepsilon_{0} \hbar}\left( \frac{1}{\omega_{jg}+\omega_{q}}+\frac{1}{\omega_{je}-\omega_{q}} \right),
  \label{polarizability-offdiagonal-q} \\ 
  &&\alpha_{ge}^{(p\overline{p})}=\alpha_{eg}^{(p\overline{p})}
  =\frac{d_{gj}d_{je}}{\varepsilon_{0} \hbar}\left( \frac{1}{\omega_{jg}-\omega_{p}}+\frac{1}{\omega_{jg}-\omega_{\overline{p}}} \right)
  =\frac{d_{gj}d_{je}}{\varepsilon_{0} \hbar}\left( \frac{1}{\omega_{je}+\omega_{p}}+\frac{1}{\omega_{je}+\omega_{\overline{p}}} \right).
  \label{polarizability-offdiagonal-p}
\end{eqnarray}
In order to include relaxation effects, it is necessary to introduce the density matrix for a mixture of pure states:  
\begin{equation}
	{\rho}=\left( \begin{array}{cc} \rho_{gg} & \rho_{ge} \\ \rho_{eg} & \rho_{ee} \end{array} \right).
        \label{density matrix}
\end{equation}
The equation of motion for the density matrix is governed by $i \hbar ({d{\rho}}/{dt})=[{H}, {\rho}]+(\mbox{relaxation terms})$, 
and its explicit forms will be shown below.
So far we have considered a single molecule, which is now extended to 
an ensemble of molecules within a finite volume.
To this end, the density matrix is considered to be a function of the position $x$ by taking a continuous limit
	of atom distribution in the target.
We also need to consider a propagation effect of the electromagnetic fields: 
this effect is included by the one-dimensional Maxwell equation
\begin{equation}
	\frac{\partial^2 E}{\partial t^2}-c^{2}\frac{\partial^2 E}{\partial x^2}=-\frac{n}{\varepsilon_{0}}\frac{\partial^2 P}{\partial t^2},
	\label{Maxwell wave equation}
\end{equation}
where $P$ denotes the macroscopic polarization, and $n$ the number density of the hydrogen molecules.
The polarization $P$ can be calculated with $P=Tr ({{\rho} d})$.   
Putting $P$ into eq.(\ref{Maxwell wave equation}) with the help of RWA and SVEA, we arrive at a set of equations, referred to as Maxwell-Bloch equation, expressed by
\begin{eqnarray}
  && \frac{\partial \rho_{gg}}{\partial \tau}=i\Big( \Omega_{ge}\rho_{eg}-\Omega_{eg}\rho_{ge} \Big)+\gamma_{1}\rho_{gg},
  \label{Bloch-gg} \\ 
  && \frac{\partial \rho_{ee}}{\partial \tau}=i\Big( \Omega_{eg}\rho_{ge}-\Omega_{ge}\rho_{eg} \Big)-\gamma_{1}\rho_{ee},
  \label{Bloch-ee} \\ 
  && \frac{\partial \rho_{ge}}{\partial \tau}=i \Big(\Omega_{gg}-\Omega_{ee}+\delta \Big)\rho_{ge}+i\Omega_{ge}\Big(\rho_{ee}-\rho_{gg} \Big)-\gamma_{2}\rho_{ge},
  \label{Bloch-ge} \\
  &&\frac{\partial E_{q}}{\partial \xi}=\frac{i \omega_{q}n}{2c}
        \Big\{\Big(\rho_{gg} \alpha_{gg}^{(q)}+\rho_{ee} \alpha_{ee}^{(q)} \Big) {E}_{q}
        +\rho_{eg} \alpha_{eg}^{(q-1)} {E}_{q-1}+ \rho_{ge} \alpha_{ge}^{(q)} {E}_{q+1} \Big\},
  \label{Maxwell-q} \\
  &&\frac{\partial E_{p}}{\partial \xi}=\frac{i \omega_{p}n}{2c}
        \Big\{ \Big(\rho_{gg} \alpha_{gg}^{(p)}+\rho_{ee} \alpha_{ee}^{(p)} \Big) {E}_{p}
        +\rho_{eg} \alpha_{ge}^{(p\overline{p})} {E}_{\overline{p}}^{\ast} \Big\}.
   \label{Maxwell-p}
\end{eqnarray}
Here we have introduced the co-moving coordinates defined by $(\tau, \xi)=(t-x/c, x)$, and the Rabi frequencies by
\begin{eqnarray}
  &&\Omega_{aa}=\frac{1}{2 \hbar}\sum_{m=p,\overline{p},q}\frac{1}{2} \varepsilon_{0}\,\alpha_{aa}^{(m)}|E_{m}|^2  \qquad (a=g,e),
  \nonumber \\ 
  &&\Omega_{ge}=\Omega_{eg}^{\ast}=\frac{1}{2 \hbar}\left\{ 
  	\sum_{q} \frac{1}{2} \varepsilon_{0}\,\alpha_{ge}^{(q)}E_{q} E_{q+1}^{\ast}+
        \frac{1}{2} \varepsilon_{0}\,\alpha_{ge}^{(p\overline{p})}E_{p}^{\ast} E_{\overline{p}}^{\ast}
        \right\}.
\end{eqnarray}
Relaxation terms in  Bloch eqs.(\ref{Bloch-gg}-\ref{Bloch-ge}), 
	given by the terms proportional to $\gamma_{1}$ (longitudinal) and $\gamma_{2}$ (transverse), 
        are the most general form in the two-level system.

\paragraph{Simulation}
Numerical simulations are performed based upon the Maxwell-Bloch equation (\ref{Bloch-gg}-\ref{Maxwell-p}) shown above.
As indicated in Table \ref{wavelength table}, the lowest $(q=-4)$ Stokes sideband is used for the trigger field for the two-photon emission in this experiment: 
	we thus take $\overline{p}$ as one of the two-photon pair and identify it with $q=-4$.
Actually, in the Maxwell eq.(\ref{Maxwell-q}) for $q=-4$, the term $\rho_{eg} \alpha_{eg}^{(q-1)} {E}_{q-1}$ was
	replaced by $\rho_{eg} \alpha_{ge}^{(p\overline{p})}{E}_{{p}}^{\ast}$. 
As to the relaxation terms, the dominant contribution comes from $\gamma_{2}$, which is  
	taken from the experimental measurements \cite{Bischel and Dyer}.
In total, 51 intermediate states $|j\rangle$ are taken into account in the evaluation of the polarizabilities, 
	and they are then rescaled so that they agree with the measured index of refraction \cite{Huang}.
The 1+1 dimensional Maxwell-Bloch equation has an apparent shortcoming: it cannot treat any transverse effects, 
	in particular the transverse intensity variation of the input lasers or the output radiations.
In Sec. 4, we will present a practical method to circumvent this insufficiency together with the simulation results.

\section{Experimental Setup}  
\begin{figure}[h]
	\begin{center}
		\includegraphics[width=0.8\textwidth]{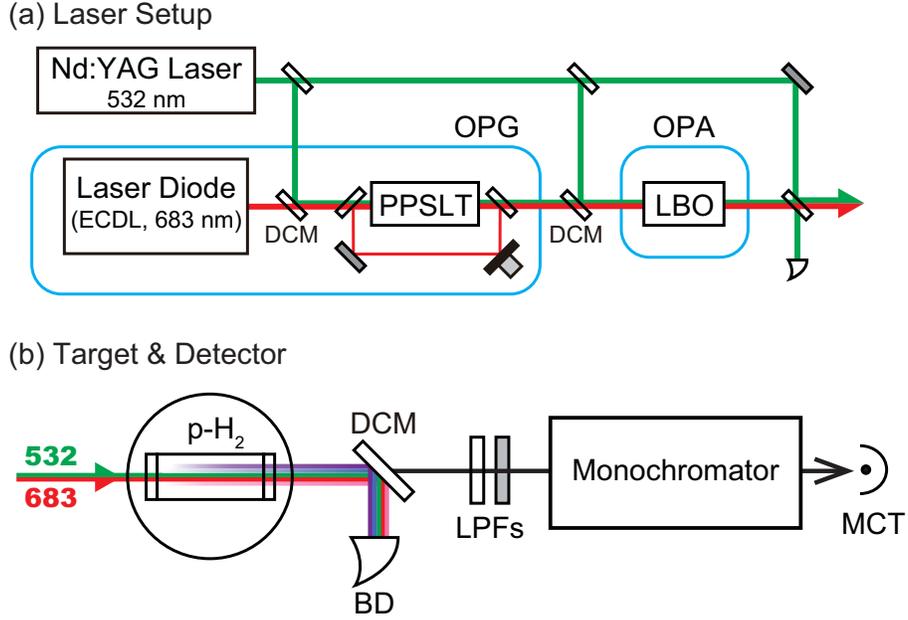}
        \end{center}
\caption{Schematics of the experimental setup. (a) The laser
 system. The main Nd:YAG laser beam is divided into three beams. 
 Two of them are used as pumping light sources to generate the $\omega_{-1}$ laser (683~nm) 
 and the rest is used as the $\omega_{0}$ laser (532~nm).
 For the $\omega_{-1}$ light generation, we employed an injection seeded OPG
 with a PPSLT crystal and OPA with LBO crystals.
 A typical output power at OPA stage is  $\ge$6~mJ at 683~nm.
 (b) Schematic diagram of the target and the detector. 
 DCM: dichroic mirror; BD: Beam dumper; LPFs: long-pass
 filters; MCT: Hg-Cd-Te mid-infrared detector.
 }
	\label{fig:Setup}
\end{figure}

Schematic diagram of the experimental setup is shown in Fig.\ref{fig:Setup}.
It consists of three major parts: the laser excitation system, the para-hydrogen (p-H$_{2}$) gas target, and the detector system.
In this experiment, generation of the large target coherence is the key to success, and every care was taken to enhance it both in the laser
	and the target systems.
In the following, we describe each in turn.

\subsection{Laser system}
We used the second harmonic of a Q-switched injection-seeded Nd:YAG
laser ($\lambda = 532.216$~nm, Litron LPY642) as a master light source; 
all the required lasers are produced from this single laser to 
reduce temporal jitter between the two pulses. 
It is operated at a repetition rate of 10~Hz with a pulse duration of 8~ns and an energy up to 130~mJ.
It has a single transverse mode ($M^2 < 1.1$) and a narrow linewidth ($<$100~MHz).
Its beam is divided into three as shown in Fig. \ref{fig:Setup}: 
one is delivered to the target as the $\omega_{0}$ laser, 
and the other two are used as pumping light sources for the $\omega_{-1}$ laser.

For the $\omega_{-1}$ light ($\lambda = 683.610$~nm) generation, 
we built a laser system of an injection-seeded optical parametric generator (OPG)
combined with an optical parametric amplifier (OPA). 
In the OPG stage, a nonlinear optical crystal of MgO-doped periodically-poled stoichiometric lithium
tantalate (PPSLT, Oxide Corp. Q1532-O001) is used; its dimension is 24~mm long $\times$ 1~mm thick
$\times$ 8~mm wide with a grating period of $10.3$~$\mu$m. 
As an injection seeding laser for OPG, an extended cavity diode laser (ECDL) in the Littrow configuration is made using a commercially available laser diode chip (TOPTICA LD-0685-0050-3, no anti-reflection coating).
The measured output power of the ECDL is more than 10~mW with a typical mode-hop-free scanning range of 3~GHz. 
The pumping (pulsed) and injection-seeding (continuous wave) laser lights are combined with a dichroic mirror, 
and then injected into the PPSLT crystal.
A typical output pulse energy at the OPG stage is 0.4~mJ, and a linewidth 
is 97~MHz, nearly Fourier transform limited linewidth.
For the OPA stage, we used bulk lithium triborate (LBO) crystals in a non-critical phase-matching condition.
The output pulse from the OPG is amplified to more than 6~mJ at the OPA stage.

The actual pulse energy and the beam waist size of the $\omega_{0}$ ($\omega_{-1}$) driving laser 
	is 4.3~mJ (4.3~mJ) and $0.12$~mm ($0.15$~mm), respectively. 
Both lasers are linearly polarized in the same direction. 
For the detuning ($\delta$) scan, we changed the frequency of
the ECDL seeding laser.

\subsection{Target}
We used para-hydrogen (p-H$_{2}$ with purity of $<500$ ppm ortho-hydrogen contamination) gas at the temperature of 78 K as a target.
The main reasons of using p-H$_{2}$ are that it is suited to observe two-photon emission from the E1 forbidden 
	vibrationally-excited state, and that the production technique of large coherence is well established.
In addition to these, 
	para-hydrogen has a merit of longer decoherence time over normal-hydrogen (1:3 mixture of para- and ortho-hydrogen), and 
        the low temperature (78 K) is better because 
        the decoherence time ($\gamma_{2}^{-1}$) is nearly the longest thanks to the Dicke narrowing
        \cite{Dicke-narrowing}.

The actual target,  cylindrical with 20 mm in diameter and 150 mm in length,  was installed in a cryostat.
The pressure could be varied, but in the present experiment it was fixed at 60 kPa 
(the estimated number density assuming ideal gas is $n=5.6 \times 10^{19}$ cm$^{-3}$).
Both pressure and temperature were monitored constantly during the experiment.
The estimated decoherence rate at this condition is about 130 MHz \cite{PTEP}.

\subsection{Detectors}
As shown in Fig.\ref{fig:Setup}(b), the lights exiting from the target cryostat window went through a dichroic mirror to reflect 
strong driving laser lights and a Ge filter to further reduce visible region lights.
They entered a monochromator (Princeton Instruments Acton SpectraPro SP2300) to analyze the wavelength of mid-infrared (MIR) lights.
The wavelength resolution of the monochromator, having a grating of 150 groove/mm and 4 $\mu$m blaze wavelength, 
was set to about 1 nm to observe MIR spectra while to about 50 nm in other experiments. 
An actual MIR  detector was MCT (HgCdTe, Daylight solutions HPC-2TE-100).
When the Raman sideband energy was measured, the system above was replaced with a prism and a pyroelectric energy detector (Gentec Electro-Optics QE12LP-H-MB).
The MCT signals were monitored by an oscilloscope and were sent to a computer for later offline analysis.
On average 100--200 shots were accumulated at a single parameter setting.

\section{Experimental Results}  
\subsection{Raman Sidebands}
\begin{figure}[t]
	\begin{center}
		\includegraphics[width=0.8\textwidth]{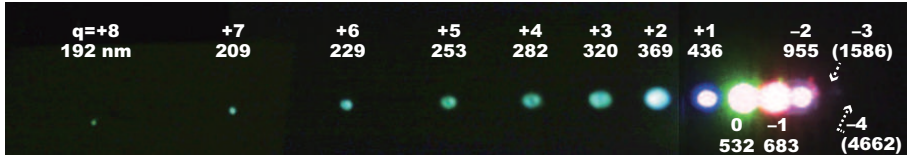}
        \end{center}
	\caption{Photograph of the Raman sidebands (projected onto a fluorescent sheet and taken by a CCD camera). 
        The wavelengths calculated with eq.(\ref{Raman sideband frequency}) are also shown.
        The third and fourth Stokes sidebands shown in parentheses are observed only by the pyroelectric energy and/or MCT detector.
        The photograph contrast and light level from $q=2$ to $q=8$ are enhanced for clear view.
        Apparent variation in the spot sizes is due to over exposure while distortion from the straight line (around $q=$6--8)
        is caused by bent of the fluorescent sheet.
        }
	\label{fig:Raman-SideBand}
\end{figure}

\begin{figure}[htb]
	\begin{center}
		\includegraphics[width=0.6\textwidth]{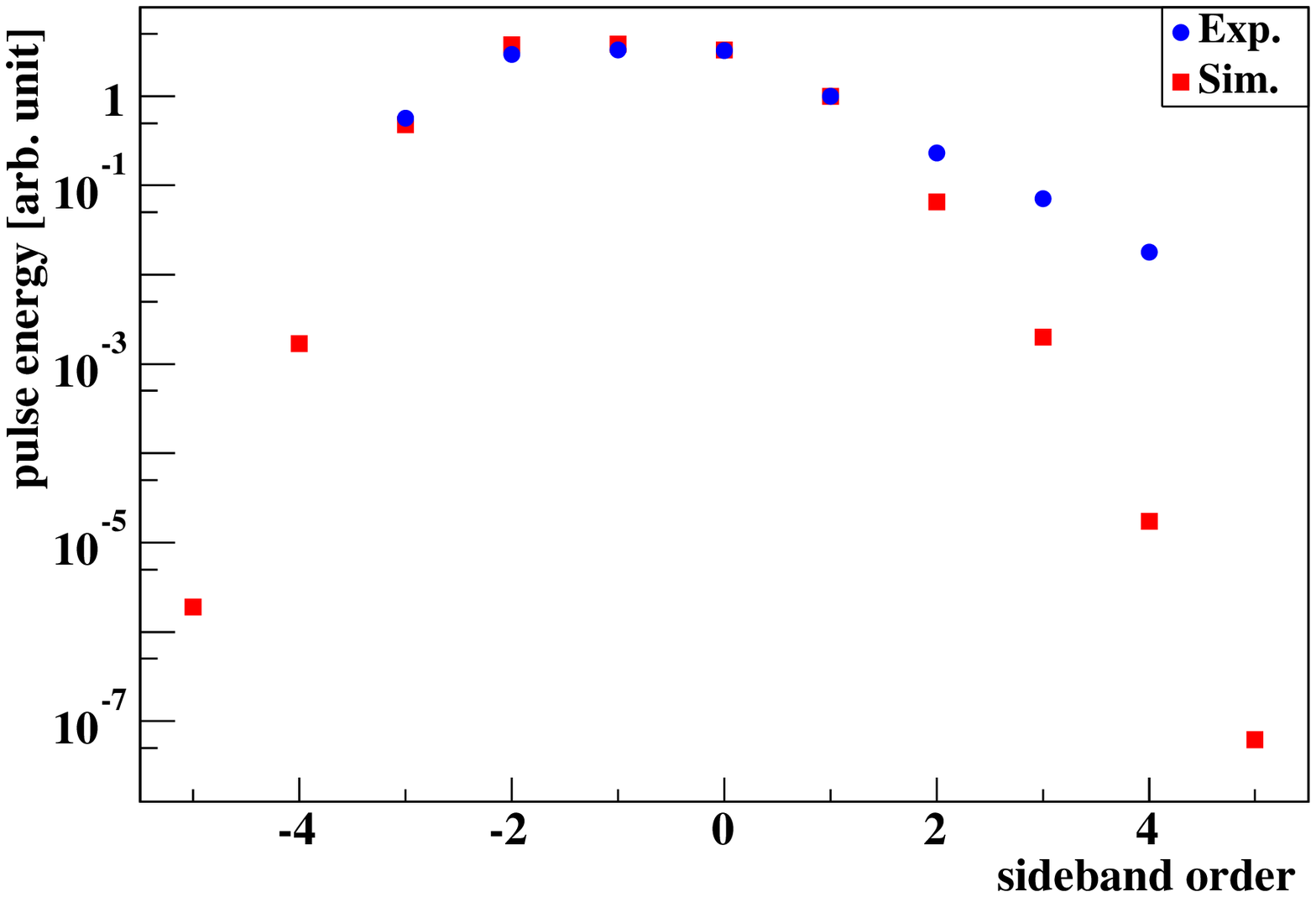}
        \end{center}
	\caption{Comparison of the Raman sideband pulse energy measurements (from $q=-3$ to  $q=+4$ at $\delta=0$) with the simulation results.
        The vertical axis represents energies (the simulation results are normalized at $q=1$) while the horizontal axis is the Raman order $q$.
         The 4.96 $\mu$m signal is plotted at $q=-5$ for convenience.
        The circles in blue (squares in red) indicate the experimental (simulation) results.}
	\label{fig:Comparison-Raman-Simulation}
\end{figure}

We first show the results of Raman sideband measurements. 
Figure \ref{fig:Raman-SideBand} shows the photograph of Raman sidebands taken by a CCD camera.
As seen, we observed the anti-Stokes sidebands up to eighth order and the Stokes sidebands to second order. 
In the photograph, the short wavelength was limited by the absorption due to the air, 
and the long wavelength by sensitivity of the CCD camera.
They were all found to be collinear with the excitation lasers.
Pulse energies of sidebands from $q=-3$ to  $q=+4$ were measured by the pyroelectric energy detector.
Figure \ref{fig:Comparison-Raman-Simulation} shows the comparison of the pulse energy measurements ($\delta=0$) with the simulation results. 
The latter is obtained as follows.
As explained in Sec.2, the 1+1 dimensional Maxwell-Bloch equation cannot handle transverse intensity variations.
This fact demands that, in the simulation, it is necessary to use an averaged power to account for the transverses intensity variation of the excitation lasers.
It also means that any radiation power obtained by the simulation must be multiplied by an angular variation factor
 (usually unknown) to compare with actual output energy measurements.   
For the input laser power, we let it free in the simulation, and determine it by seeking the best fit to the actual data\cite{chi2}.
For the output power, we obtain the needed angular factor using one of the sideband data, say $q=1$.
In other words, all the simulation results (including the 4.96 $\mu$m emission) are multiplied by a common factor 
	so that the $q=1$ sideband pulse energy agrees with the corresponding experimental result.
Actually, the best input power in the simulation is found to be about 0.39  (0.14) of the peak power for the $\omega_{0}$ ($\omega_{-1}$) laser\cite{chi2}.
As seen in Fig.\ref{fig:Comparison-Raman-Simulation}, the overall agreement between the simulation and experimental results is satisfactory,
although the simulation predicts lower power for large $q$, say $q>3$. 
From this simulation result, we can estimate an average degree of coherence along the target:  
it is $\rho_{ge}\simeq 0.032$ at $\tau=0$, the peak timing of the driving lasers.

\subsection{Two-photon emission process}
\begin{figure}[htb]
	\begin{center}
		\includegraphics[width=0.7\textwidth]{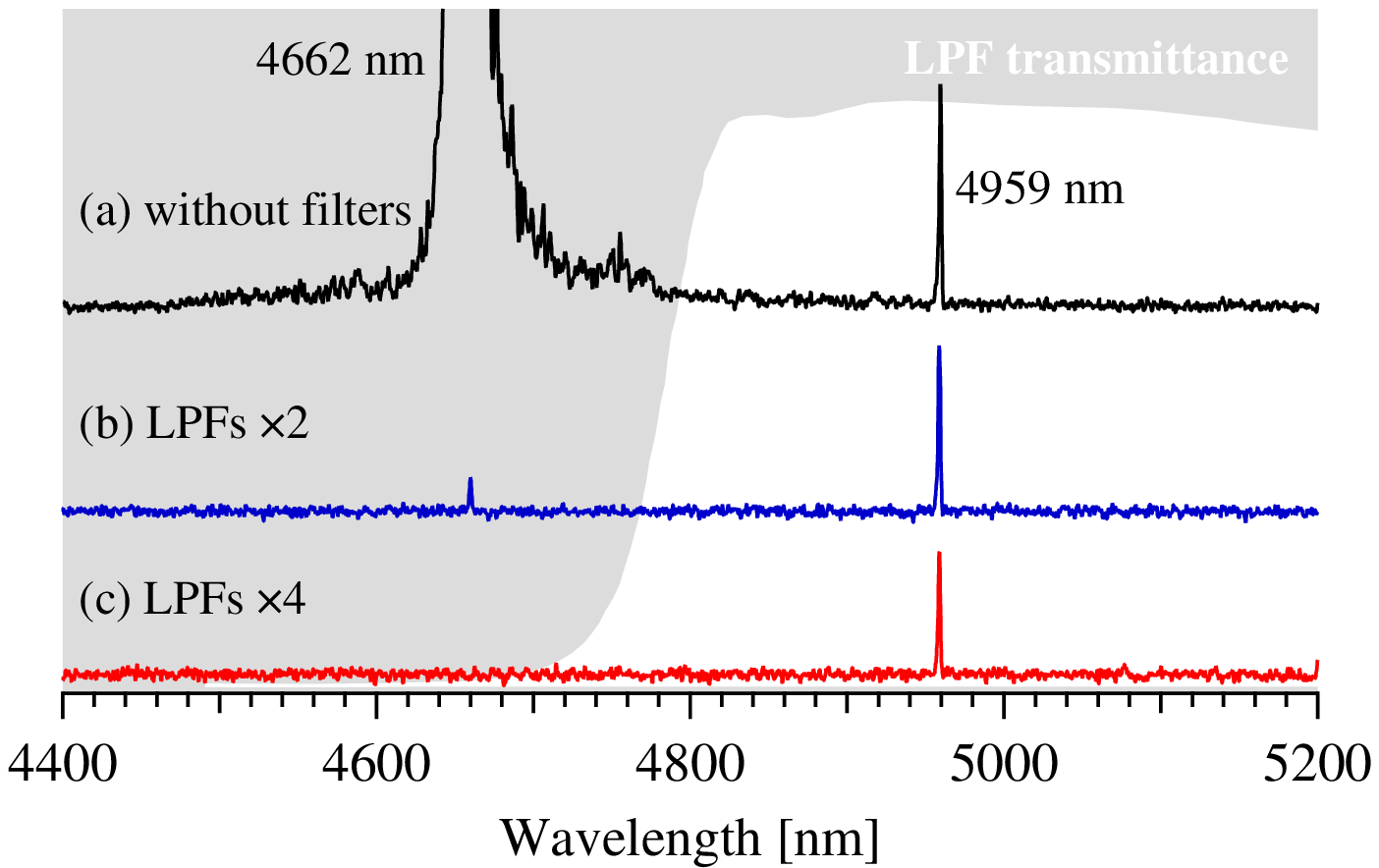}
        \end{center}
	\caption{Observed spectra at $\delta=0$ MHz and 60 kPa ; (a) without the longpass filter (LPF), (b) with two LPFs, and (c) with four LPFs. 
        The white portion excluded by the gray hatch shows the LPF transmittance; it is $\sim$0.85 at 4.96 $\mu$m.}
	\label{fig: MIR-Spectrum-Result}
\end{figure}

Figure \ref{fig: MIR-Spectrum-Result} shows the result of spectrum measurements at the detuning of $\delta=0$.
The black line is the spectrum without the long-pass filter (LPF, Spectrogon LP-4700nm)
   while the blue (red) line is the one with two (four) LPFs inserted
   in front of the monochromator.
The transmittance of the LPF is indicated by the white portion excluded by the gray hatch.
Two peaks were unambiguously observed corresponding to the fourth Stokes sideband (4.66 $\mu$m) and its two-photon partner (4.96 $\mu$m).
The 4.66 $\mu$m signal saturated the detector without LPF, but was mostly filtered out with LPFs.
On the other hand, the 4.96 $\mu$m signal remained unaffected with and without LPFs 
(the peak heights reduced by LPF transmittance of $\sim 0.85$ per a filter): 
this fact eliminates the possibility of a higher order reflection light of the grating system.
It was found that these signals had a sharp forward distribution 
(half angular divergence of $\sim$20 mrad for 4.66 $\mu$m and $\sim$10 mrad for 4.96 $\mu$m)
and a time profile similar to the input driving lasers 
(with slightly narrower FWHM pulse durations of ~5 ns).
The latter can be interpreted as a measure of the duration time of the produced coherence.

The ratio of the two signals, defined by the 4.96 $\mu$m energies divided by those of 4.66 $\mu$m,  
is $\sim0.8 \times 10^{-3}$ at this detuning.

\begin{figure}[htb]
	\begin{center}
		\includegraphics[width=0.5\textwidth]{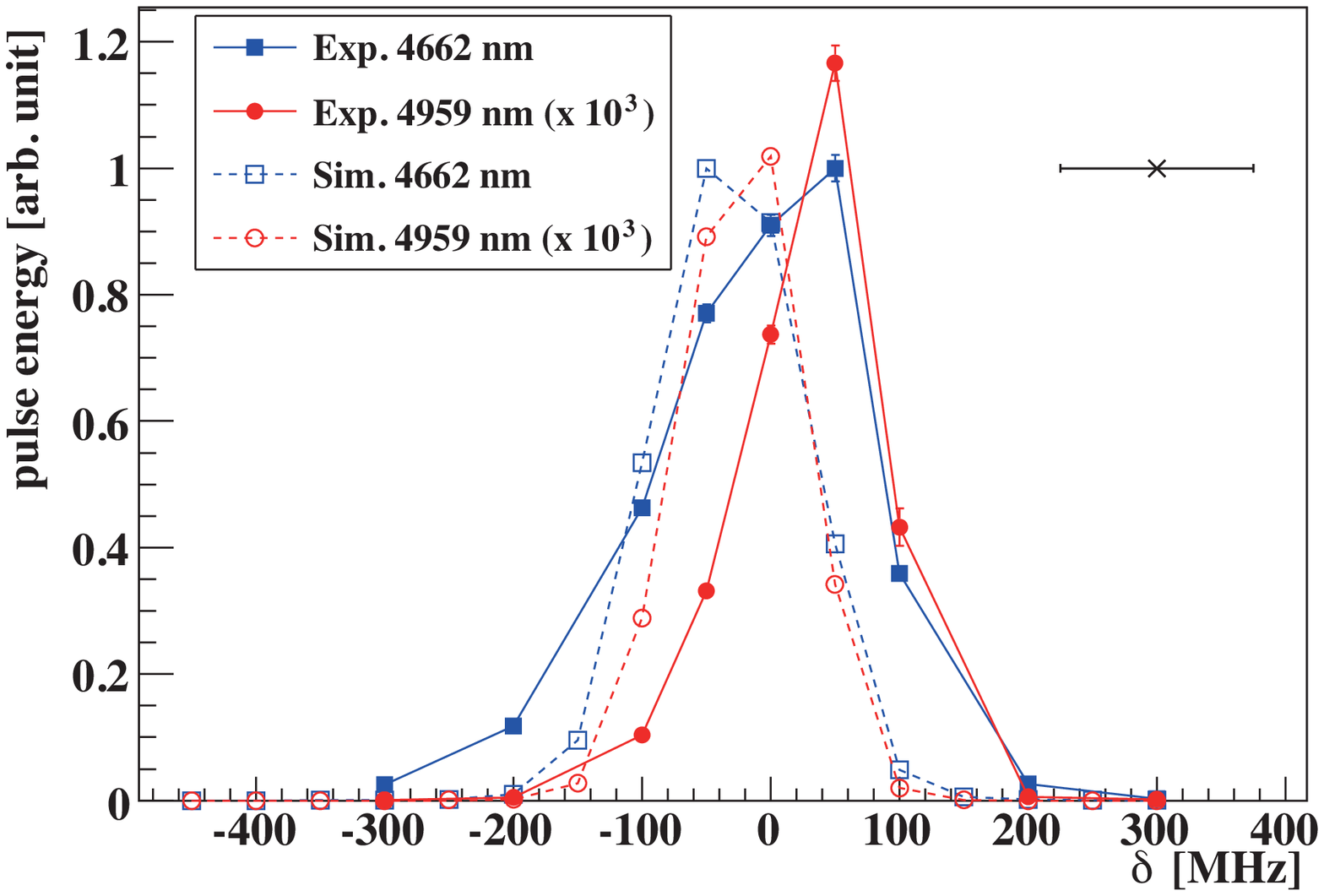}
        \end{center}
	\label{fig:Detuning-Curve-all}
	\caption{The 4.66 and 4.96 $\mu$m output pulse energies as a function of the detuning frequency $\delta$.
        The solid (open) symbols connected by solid (dashed) lines indicate the experimental (simulation) data. 
        The red circles are for 4.96 $\mu$m (scaled up  by $10^{3}$) and the blue squares for 4.66 $\mu$m. 
        The horizontal bar in the plot indicates $\pm 75$ MHz uncertainty in the frequency measurements.
        }
\end{figure}

\subsection{Detuning curve}
Figure \ref{fig:Detuning-Curve-all} shows pulse energies of the $4.66$ and $4.96$ $\mu$m pair 
	as a function of the detuning $\delta$ (detuning curves). 
In the figure, the experimental data (indicated by solid squares and circles) are obtained by integrating each MCT output pulse while 
	the simulation data (indicated by open squares and circles) are normalized in such a way that 
        the maximum values of the 4.66 $\mu$m real and simulation data agree with each other.
Thus meaningful comparisons between the real data and simulations are the shape of the $4.66$ and $4.96$ $\mu$m detuning curves
	and their relative magnitude.
As to the shape,  the agreement between the real data and simulations is good.
However, the peak positions  of the real data (for both 4.66 and 4.96 $\mu$m) are slightly ($\sim100$ MHz) higher than those of the simulation data.
We note that absolute accuracy of the frequency determination is estimated to be $\pm 75$ MHz\cite{FrequencyAccuracy}; 
thus the difference in the peak positions may stem from the uncertainty in the frequency measurements.
As to the relative magnitude, the ratios of 4.96~$\mu$m to 4.66~$\mu$m powers are 
	in the order of $10^{-3}$ both for the real and simulation data,
	showing good agreement each other.
In any case, the overall agreement between the simulation and experimental results is regarded reasonable.

For illustrative purpose, we compare below the $4.96$ $\mu$m absolute pulse energy with the spontaneous two-photon decay.
To this end, the measured outputs are corrected for various transmittance or reflection efficiencies of the optical elements, 
except for the monochromator efficiency which is assumed to be unity\cite{Monochromator-Acceptance}. 
The resulted value, converted to the number of photons per pulse, is $4.4 \times 10^{7}$. 
As to the spontaneous decay process, we have estimated it as follows.
Its rate ($A$) is expressed by 
\begin{equation}
	\frac{dA}{dz}=\frac{\omega_{eg}^{7}}{(2\pi)^3 c^{6}} \left| \alpha_{ge}^{(p\overline{p})}\right|^{2} z^{3}(1-z)^{3}
        \sim 3.2 \times 10^{-11} \mbox{ 1/s} \quad (z=\frac{1}{2})
\end{equation}
where $z=\omega/\omega_{eg}$ is the fractional energy of one of the two photons.
Considering the energy band width $\Delta z \sim 4.9 \times 10^{-3}$ (taken to be the monochromator full width), 
the measurement time $\Delta t \sim 80$ [ns],  
the detector solid angle fraction $\Delta \Omega/(4\pi)\sim 1.2 \times 10^{-4}$ (for which the monochromator efficiency is assumed to be unity),  
and the maximum number of excited states in the target ($\sim 1.5 \times 10^{16}$), 
we obtained the number of expected photons to be $1.6 \times 10^{-8}$ per pulse.
This value may be compared to $4.4 \times 10^{7}$, which is lower bound of the photons actually observed: 
the huge enhancement factor ($> 10^{15}$) can only be understood in the presence of macro-coherence.

\section{Conclusions and Summary}
In this paper, we have described an experiment which was conducted to explore the macro-coherent amplification mechanism 
using the two-photon emission process from the para-hydrogen electronically-ground vibrationally-excited state ($Xv=1$).
The adiabatic Raman method was employed to prepare large coherence in the initial state.
The Raman sidebands from the lowest Stokes ($q=-4$) up to the eighth anti-Stokes ($q=8$) were observed and 
	their intensities were compared to the simulation 
	based upon the Maxwell-Bloch equation in order to estimate the degree of coherence.
With the lowest Stokes sideband ($\lambda =4.66\ \mu$m) used as a trigger, 
	the two-photon emission partner ($\lambda =4.96\ \mu$m) was seen unambiguously.
The observed two-photon rate is found to be much larger than that of the two-photon spontaneous decay, 
	and to be consistent with the expectation of the Maxwell-Bloch equation derived for the process.
Although the macro-coherence amplification mechanism deserves further examination, 
the present experimental results support its basic principle in the non-explosive regime.

\newpage
\section*{Acknowledgment}
We thank Professors K. Kawaguchi and T. Momose for valuable discussions.
We are deeply indebted to Prof. M. Katsuragawa for his advices on the adiabatic Raman process. 
This research was partially supported by   
Grant-in-Aid for Scientific Research on Innovative Areas "Extreme quantum world opened up by atoms"
(21104002), Grant-in-Aid for Scientific Research A (21244032), 
Grant-in-Aid for Scientific Research C (25400257), 
Grant-in-Aid for Challenging Exploratory Research (24654132),
and Grant-in-Aid for Young Scientists B (25820144)
from the Ministry of Education, Culture, Sports, Science, and Technology.

\vspace{1cm}
$\;$\\


\begin{thebibliography}{9}
\bibitem{PDC2}
The Bloch equation, not  explicitly written in Ref.\cite{Parametric-Down-Conversion}, 
is necessary to take into account detailed behavior of atoms.
Comparison of experimental data with numerical simulations is done 
based on the Maxwell-Bloch equation in this paper.
\bibitem{polarizability-note}
 The polarizability $\alpha_{gg}$ is related to the index of refraction by  $n_{r} \simeq 1+ n\alpha_{gg}/2$ in the long wavelength limit, 
 where $n$ is the number density of the molecules.
\bibitem{chi2}
We note that the Raman sidebands are generated only from the overlapped volume of the two driving lasers: 
	however, they do not overlap completely in general.
Taking this possibility into account, we treated their effective intensities independent,  and  
 minimized the $\chi^2$ defined by 
\begin{equation}
	\chi^2=\sum_{q \ne 0, -1}  
        \left( \frac{P_{q}^{\ \mbox{exp}}(\delta=0)-P_{q}^{\ \mbox{sim}}(\delta=0) } {P_{q}^{\ \mbox{exp}}(\delta=0)}\right)^2
\end{equation}
where $P_{q}^{\ \mbox{exp}}(\delta=0)$ ($P_{q}^{\ \mbox{sim}}(\delta=0)$) denotes 
the $q$-th sideband power normalized at $q=1$ obtained by the experiment (simulation) at the detuning $\delta=0$.
The summation runs from $q=-3$  to $q=3$ excluding $q=0$ and $q=-1$. 
These powers, corresponding to the two driving lasers ($\omega_{0}$ and $\omega_{-1}$), should be excluded from the $\chi^2$ sum because 
they may contain ineffective ones (non-overlapping part).
\bibitem{Monochromator-Acceptance}
The monochromator efficiency depends on the spot size and angular divergence of input lights 
and/or grating efficiency which in turn depends on the wavelength.
It is difficult to measure this efficiency accurately without a good MIR calibration light source.
The number of photons quoted in the paper thus represents the lower bound.
\end{thebibliography}
\end{document}